\documentclass[twocolumn,prl,floatfix,showpacs]{revtex4}
\usepackage{color}
\usepackage{epsfig}
\usepackage{graphicx}
\usepackage{amsmath,amssymb}
\newcommand{\p}{\partial}
\newcommand{\e}{\epsilon}
\newcommand{\s}{\sigma}
\newcommand{\m}{\mu}
\newcommand{\g}{\gamma}
\renewcommand{\b}{\beta}

\renewcommand{\a}{\alpha}

\renewcommand{\v}{\varepsilon}
\newcommand{\D}{\Delta}

\begin{document}

\title{ Theory of dendritic growth in the presence of lattice strain.}

\author{D. Pilipenko}
\author{E. A. Brener}
\author{C. H\"uter}
\affiliation{Institut f\"ur Festk\"orperforschung, 
Forschungszentrum J\"ulich, D-52425 J\"ulich, Germany}


\date{\today}

\begin{abstract}
 Elastic effects due to lattice strain modify the local equilibrium condition 
at the solid-solid interface compared to the classical dendritic growth.
Both, the thermal and the  elastic fields are  eliminated by the Green's function techniques 
and a closed nonlinear integro-differential equation for the evolution of the 
interface is derived.  In the case of  pure dilatation, the 
elastic effects lead only to a trivial shift of the transition temperature while
in the case of shear transitions, 
 dendritic patterns are found  even for isotropic  surface energy.

\end{abstract}
\maketitle
Solvability theory has been very successful in predicting
certain properties of dendritic growth and a number of related phenomena 
(see, for example, \cite{kessler,brenermelnikov}). 
The solution of  the two-dimensional steady state growth is described by 
a needle crystal, which  is assumed to be close
in shape to the parabolic Ivantsov solution \cite{Ivantsov}.
If anisotropic capillary effects are included, a single dynamically stable
solution is found for any external growth condition. This
theory has also been extended to the three-dimensional case
\cite{brener93,benamar93}. 
The capillarity is a singular perturbation
and the anisotropy of the surface energy is a prerequisite
for the existence of the solution.
In the case of isotropic surface energy, 
dendritic solution does not exist and instead the so-called doublon 
structure is the solution of the problem \cite{benamar95,ihle93}.

Usually, the structural phase transitions in solids are accompanied by small lattice 
distortions leading to elastic deformations (for a review, see \cite{roitburd74, 
khachaturyan83}).
One of the well-known consequences is a thermodynamic
elastic hysteresis-i.e., the splitting of the phase equilibrium
point into two points: the point of the direct and the inverse
transition. This is mainly due to the coherency at the interface
boundary, meaning that the lattice layers remain continuous
through the boundary. Correspondingly, the hysteresis disappears
without interface coherence \cite{brener92,brener99}. However, the systematic investigation of 
the growth {\it kinetics} of such phase transitions is much less developed. Recently, 
the pattern formation processes controlled by interface kinetics have been considered 
in \cite{brenerspatschek} and the growth of spherical inclusions 
under elastic and thermal influence were investigated 
in \cite{slutsker06} by means of the phase-field model.  

In this Letter, we discuss the influence of elastic strain on dendritic growth in 
solids   controlled by heat diffusion.
Significant progress in the description of  dendritic growth was made by the 
elimination of the thermal field
 using the Green's function technique. This allows to obtain a closed  equation  
for the interface evolution (see for example \cite{langer80}).
The crucial point of the present analysis is that the elastic field can also be  
eliminated by the corresponding Green's function technique.
By these means, we  derive here, as in the classical dendritic growth theory, a 
single integro-differential equation for the shape of the interface which takes into 
account elastic effects.  
 Then, we consider two simple examples of  dilatational and shear transformations. 
We show that in the case of a pure dilatation,  the elastic effects lead only to 
a trivial shift of the transition temperature. 
However, for the  case of shear transitions, we find dendritic patterns 
even without the anisotropy of the surface energy required by the classical dendritic
growth  theory. 
In this sense, the elastic effects serve as a new selection mechanism.
%%%%%%%%%%%%%%%%%%%%%%%%%%%%%%%%%%%%%%%%%%%%%%%%%%%%%%%%%%%%%%%%%%%%%%%%%%%%%%%%%%%%%%%%%%%%%%%%%%%%%%%%%%%%%
%%%%%%%%%%%%%%%%%%%%%%%%%%%%%%%%%%%%%%%%%%%%%%%%%%%%%%%%%%%%%%%%%%%%%%%%%%%%%%%%%%%%%%%%%%%%%%%%%%%%%%%%%%%%%  

{ \em Thermodynamics of the model.}
Let us consider the growth of a new $\beta$-phase inside of an unbounded mother $\a$-phase.
We denote the characteristic lattice strain (also known  as the stress-free strain tensor), 
associated with
the phase transition, by $\e_{ik}^{0}$. 
The  free energy density in the initial  $\a$-phase is   
\begin{eqnarray}
 F_{\a}=F_{\a}^0(T)+\frac{\lambda}{2}\e_{ii}^2+\m \e_{ik}^2,
\label{fr1}
\end{eqnarray} 
where $F_{\a}^0(T)$ is the free energy density  without elastic effects, 
which depends only on the temperature $T$,
$\e_{ik}$ are the components of the strain tensor, and $\lambda$ and $\m$ are the 
elastic moduli of isotropic linear elasticity. The  free energy density of the growing 
$\b$-phase is given by:
\begin{eqnarray}
  F_{\beta}=F_{\beta}^0(T)+\frac{\lambda}{2}\left(\e_{ii}-\e_{ii}^{0}\right)^2+\m\left(\e_{ik}-\e_{ik}^{0}\right)^2.
 \label{fr2}
 \end{eqnarray}
Here, we neglect the difference between the elastic coefficients in the two phases. 
We also assume that the elastic effects are small i.e., $\e^{0}_{ik}\ll 1$.
Since in our description the reference state for both phases is the undeformed initial phase, the coherency condition at the interface reads $u_i^{(\a)}=u_{i}^{(\b)}$, where $u_i$ is the displacement vector. 
Mechanical equilibrium at the interface demands $\s_{nn}^{(\a)}=\s_{nn}^{(\b)}$ and $\s_{n\tau}^{(\a)}=\s_{n\tau}^{(\b)}$, $\s_{ns}^{(\a)}=\s_{ns}^{(\b)}$.
Here indices $n$  and ($\tau$, $s$) denote the normal and tangential directions 
with respect to the interface; the stress tensor is defined as: 
\begin{eqnarray*}
 \s_{ik}=\frac{1}{2}\left(\frac{\p F}{\p \e_{ik}}+\frac{\p F}{\p \e_{ki}} \right).
\end{eqnarray*} 

The  condition of phase equilibrium requires the continuity of a 
new potential
\begin{eqnarray*}
 \tilde{F}=F-\sigma_{nn}\e_{nn}-2\sigma_{n\tau}\e_{n\tau}-2\sigma_{ns}\e_{ns}
 \end{eqnarray*}
across the flat interface \cite{privototsky}, which takes into account the
coherency constraint.
In the general case of curved interfaces, the surface energy $\g$ has also to be 
incorporated, and the
phase equilibrium condition for each interface point in
the case of isotropic surface energy reads
\begin{eqnarray}
 \tilde{F}_{\a}-\tilde{F}_{\b}-\g \kappa=0,
\label{local}
\end{eqnarray}
where $\kappa$ is the local curvature of the interface. 

{\em Solution of the elastic problem.} 
Let us denote by $\tilde{\s}_{ik}$ the stress tensor which is related to the strain field $\e_{ik}$ by the usual Hooke's law.
Then, the stress in the new  $\b$-phase can be written as 
$\s^{(\b)}_{ik}=\tilde{\s}^{(\b)}_{ik}-\s^{0}_{ik}$, while it remains unchanged in the mother $\a$-phase: 
$\s^{(\a)}_{ik}=\tilde{\s}^{(\a)}_{ik}$. 
The tensor $\s^{0}_{ik}$ is related to  the lattice strain $\e_{ik}^0$ by Hooke's law
\begin{eqnarray}
\s^{0}_{ik}=\frac{E}{1+\nu}\left(\e^{0}_{ik}+\frac{\nu}{1-2\nu}\delta_{ik}\e^{0}_{ll}\right),
\label{sigma}
\end{eqnarray}
where  $E$ is the Young's modulus and $\nu$ is the Poisson ratio. Thus, the mechanical 
equilibrium conditions at the interface require us to introduce the interface force density in the 
equilibrium  equation $\partial\tilde\sigma_{ik}/\partial x_k=f_i$,
\begin{eqnarray}
 f_i=(\tilde{\s}^{(\b)}_{ik}-\tilde{\s}^{(\a)}_{ik})n_k =\s^{0}_{ik}n_k,
\label{force0}
\end{eqnarray} 
where   the normal vector ${\bf n}$ points from the  
$\b$-phase into the  $\a$-phase.
Because the forces act only at the interface, the displacement field 
can be written as an integral over the interface surface:
\begin{eqnarray*}
 u_i({\bf r})=\int G_{ik}({\bf r,r'}) f_k({\bf r'}) dS',
\end{eqnarray*}
where $G_{ik}({\bf r,r'})$ is the so called Green's tensor (see, for example, 
\cite{landau}).
Then, the strain field is determined by:
\begin{eqnarray}
\e_{ik}({\bf r})=\frac{1}{2}\int \left(\frac{\p G_{km}({\bf r,r'})}{\p x_i}+\frac{\p 
G_{im}({\bf r,r'})}{\p x_k} \right) f_m({\bf r'})dS'. 
\label{strain}
\end{eqnarray}
This strain is fully defined by the corresponding Green's tensor and Eqs.~(\ref{sigma},\ref{force0}).
The strain components $\e_{\tau\tau},\e_{s\tau},\e_{ss}$ are continuous across the 
interface while the other components are discontinuous:
\begin{align*}
\e_{nn}^{(\b)}-\e_{nn}^{(\a)}&=\e^{0}_{nn}+\frac{\nu}{1-\nu}\left( \e^{0}_{\tau\tau}+\e^{0}_{ss}\right),\\ 
\e_{n\tau}^{(\b)}-\e_{n\tau}^{(\a)}&=\e^{0}_{n\tau},\quad \e_{ns}^{(\b)}-\e_{ns}^{(\a)}=\e^{0}_{ns}.
\end{align*}
Taking  these jumps of the strain field at the interface into account, one can  find 
the elastic contribution to the local equilibrium condition, Eq. (\ref{local}). 
A tedious but  straightforward calculation leads to:  
\begin{align}
\label{df}
\delta\tilde{F}^{(el)}=\tilde{F}^{(el)}_{\a}-\tilde{F}^{(el)}_{\b}=\s^{(0)}_{ik} 
\e_{ik}^{(\a)}-\\ \nonumber
 -\frac{E[(\e^{0}_{\tau\tau})^2+(\e^0_{ss})^2+2\nu\e^0_{ss}\e^0_{\tau\tau} +
2(1-\nu)(\e^{0}_{s\tau})^2]}{2(1-\nu^2)},
 \end{align}
where $\e_{ik}^{(\a)}$ is the strain in the $\alpha$ phase at the interface.  
Note that the expression above  is a complicated integro-differential  functional of 
the interface shape.

\begin{figure}
\begin{center}
\epsfig{file=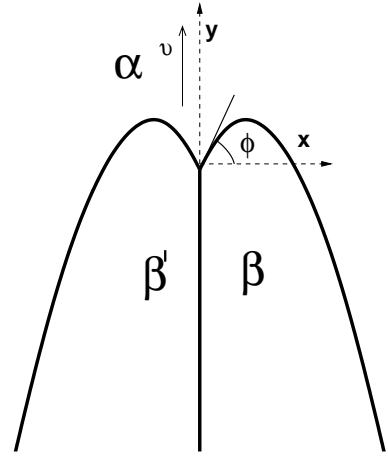, angle=0, width=5cm}
\caption{Steady state growth of a bicrystal  $(\b/\b')$.
The structure propagates with a constant velocity $v$ along the $y$ axis. 
 The $\b$-phase corresponds to $\theta=2\pi/3$ and  the $\b'$-phase corresponds to 
$\theta=-2\pi/3$.}
\label{boundary}
\end{center}
\end{figure}
 
{ \em Diffusional growth.}
For simplicity, we consider
transitions in pure materials and assume that the heat diffusion constants are equal in 
both phases (the so-called symmetrical model). 
Without loss of generality, we assume that the  $\b$ phase is the low temperature phase.
We  introduce  the dimensionless temperature field $w=c_p(T-T_{\infty})/L$, where 
$L$ is the latent heat,  $c_p$ is heat capacity, and $T_\infty$ is the temperature 
in the  $\a$ phase far away from the interface.  
The temperature field $w$ obeys the following heat diffusion equation and boundary 
conditions
\begin{eqnarray}
D\nabla^2 w=\partial w/\partial t,\label{heatdiff}\\
\upsilon_n= D{\bf {n}}({\bf{\nabla}} w_{\b} |_{int}-{\bf{\nabla}} w_{\a} |_{int}),\label{energy}\\
 w|_{int}=\D -d_0\kappa +T_{eq}c_p\delta\tilde{F}^{el}/L^2,\label{localeq}
\end{eqnarray}
where $d_0 = \gamma T_{eq}c_p/L^2$ is the capillarity length, $\kappa$ is the curvature of the interface, which assumed to be positive for convex interfaces, $D$ is the thermal diffusion constant, and
$T_{eq}$ is the equilibrium temperature for the flat interface  
without elastic effects i.e, 
it is determined by the condition $F_{\a}^0(T_{eq})=F_{\b}^0(T_{eq})$.
We also introduce the dimensionless undercooling  $\Delta = c_p(T_{eq}-T_{\infty})/L$. 
The physics underlying  Eqs.~(\ref{heatdiff}-\ref{localeq}) is quite simple.
The interface moving with the normal velocity $\upsilon_n$ releases the latent heat. 
Requirement of  heat conservation 
at the interface gives  Eq.~(\ref{energy}). 
The local thermodynamical equilibrium at the interface, Eq. (\ref{local}), implies 
Eq.~(\ref{localeq}), which gives the equilibrium 
value of the temperature at the interface taking into account the curvature corrections 
and the elastic effects. 
The thermal field can be eliminated  by using the  
Green's function techniques (see for example \cite{langer80}), 
and consequently together with a proper Green's tensor $G_{ik}(\bf{r,r'})$ 
for the elastic field, the set 
of Eqs.~(\ref{strain}-\ref{localeq}) can be incorporated into a single integro-differential 
equation  for the  shape of the solid-solid interface. 

{\em Dilatation.}
First, we consider the dilatational case, $\e^{0}_{ik}=\v\delta_{ik}$, where the bond 
lengths of the new phase are uniformly longer or shorter in all directions in comparison to the original phase.
In this case, the elastic contribution to the local equilibrium condition at 
the interface,
 $\delta \tilde{F}^{(el)}=-\v^2E/(1-\nu)$, is a constant along the interface for any  
interface shape. This result corresponds to the elastic hysteresis mentioned above
and it can be obtained using the analogy
of this elastic problem to the problem of thermal expansion
for a given temperature field \cite{landau}.
As a consequence, the equilibrium interface temperature is shifted by a constant 
value, and the problem is equivalent to the problem of the classical dendritic growth. The anisotropy of the surface energy is a prerequisite for the existence of the dendritic solution in this case,  as we have already mentioned above.
Note that this exact result is valid only under the assumptions of our model i.e., 
the isotropic elasticity and equal  elastic moduli in both phases. 
It also serves as a nontrivial check of our numerical code.

{\em Shear transition.} 
Let us  consider  now a simple type of  transition in  hexagonal crystals involving the 
shear strain. For the transitions lowering the symmetry from $C_6$ 
to $C_2$ the shear strain in the basic plane appears. This is the case, for example, 
in hexagonal-orthorhombic transitions in ferroelastics (see \cite{Jacobs01} and 
references therein).  Let the principal axis $C_6$ be orientated in the $z$ direction. 
We assume that the system obeys a translational invariance in this direction, 
and thus it is effectively two-dimensional. By proper choice of the crystal orientation 
around the main axis in the initial phase, we obtain the new phase in three possible 
states, having the following nonvanishing components of the strain tensor $\e^{0}_{ik}$:
\begin{eqnarray}
 \e^{0}_{xx}=-\e^{0}_{yy}=\v\cos 2\theta \quad \e^{0}_{xy}=\v\sin 2\theta,
\label{strain1}
\end{eqnarray}
where the angle $\theta$ has three possible values: $\theta=0, \pm 2\pi/3$.
The situation of $\theta=0$ corresponds to the single crystal growth and 
$\theta=\pm2\pi/3$ to the bicrystal growth (Fig.1).  
 Because the elasticity of hexagonal crystals is axisymmetric in the harmonic approximation 
and $\e^0_{iz}=\e_{iz}=0$, we can use the isotropic theory of elasticity i.e., 
expressions for free energy (\ref{fr1},\ref{fr2}) remain valid \cite{landau}. The moduli 
of the effective isotropic elasticity, $\lambda$ and $\mu$, can be expressed in terms of 
the elastic constants of the original hexagonal crystal.
The two-dimensional Green's tensor for isotropic materials is \cite{brebbia}: 
\begin{eqnarray}
 G_{ik}({\bf r})=\frac{1+\nu}{4\pi(1-\nu) E}\left(\frac{x_ix_k}{r^2}-(3-4\nu)
\delta_{ik}\ln( r)\right).
 \label{green}
\end{eqnarray}
Eliminating the thermal field, we obtain the steady state equation  for 
the shape of the solid-solid interface.  In the co-moving frame of reference, this 
equation reads: 
\begin{eqnarray}
\label{full_nonlin_main}
\D-\frac{d_0\kappa}{R}+\frac{T_{eq}c_p\delta\tilde{F}^{el}}{L^2}=\\ \nonumber
=\frac{p}{\pi}\int dx'\exp[-p(y(x)-y(x'))]K_0(p\eta),
\end{eqnarray} 
where $\eta = [(x-x')^2+(y(x)-y(x'))^2]^{\frac{1}{2}}$, and  $K_0$ is the modified 
Bessel function of third kind in zeroth order, and $p =  \upsilon R/2 D$ is the Peclet 
number. All lengths are reduced by the radius of the curvature $R$ of asymptotic 
Ivantsov parabola. 

In the asymptotic region ($|x| \rightarrow \infty$), the strain $\epsilon_{ik}$ 
decays and 
the local contribution to  $\delta\tilde{F}^{el}$ (second term in Eq.~\ref{df}) has a 
constant  value, $\delta\tilde{F}^{el}= -E(\epsilon_{yy}^0)^2/2(1-\nu^2)$.
It  follows from this relation that the temperature shift (elastic hysteresis) 
for the growth of a  single crystal  is four times larger than for a   
bicrystal. Although both bicrystal configurations ($\b/\b'$ and $\b'/\b$) 
are energetically equivalent far from the tip, 
the symmetry is broken by the choice of the propagation direction.
Therefore,  in the following we will discuss  the most  
favorable configuration of bicrystals \cite{brenerspatschek},  as presented in  
Fig.~\ref{boundary}.  

Let us introduce the shifted, due to the elastic hysteresis, undercooling:
\begin{eqnarray*} 
\tilde{\Delta}=\Delta-\Delta_{el}, \quad \Delta_{el}=T_{eq}c_pE\v^2/8(1-\nu^2)L^2.
\end{eqnarray*} 
The dimensionless parameter $\Delta_{el}$ describes the strength of the elastic effects.
The relation between this shifted undercooling $\tilde\D$ and the Peclet number 
is given by the  two-dimensional Ivantsov formula \cite{Ivantsov}: $\tilde\D 
= \sqrt{p \pi} \exp{(p)} \text{ erfc}(\sqrt{p})$.

The presence of the twin ($\b/\b'$) boundary leads to additional effects. First,  
calculating the strain field, $\epsilon_{ik}$, which enters in the expression for 
$\delta\tilde{F}^{(el)}$, in addition to the integrals 
along $\a/\b$ and $\a/\b'$ interface, the integration should also be performed along the 
twin boundary $\b/\b'$. The force density at this boundary is: $f_x=0, 
f_y=E \sqrt{3}/(1+\nu)$. Second, the equilibrium angle $\phi$ at the triple junction 
 (Fig. 1) is given by 
Young's law: $2 \gamma \sin \phi = \gamma_b$, where $\gamma$ is the surface 
energy of the $\b/\a$ interface  and $\gamma_b$ is the surface energy of 
the twin boundary.

 Eq. (\ref{full_nonlin_main}) is a complicated nonlinear integro-differential equation
for the interface shape. We should find a solution of this equation
which has a proper angle $\phi$ at the triple junction and which is close to the 
Ivantsov parabola ($y=-x^2/2$) in the tail region. 
Note that without elastic effects, this problem  is equivalent to the classical  
dendritic growth problem with isotropic surface tension. 
The latter does not have a solution with  angles $\phi \ge 0$ 
\cite{kessler,brenermelnikov}. This statement can be expressed in the following form. 
For any given positive values of the Peclet number $p$ 
and the so-called stability parameter 
$\sigma=d_0/pR$, the symmetric solution which is close to the Ivantsov parabola in the 
tail region has an angle at the tip $\phi=f(\sigma,p)<0$. The limit $\sigma=0$ and 
$\phi=0$ is a singular limit for that problem.  
For example, Meiron \cite{meiron}  calculated the angle $\phi$ as a function
of $\sigma$ for several values of the Peclet number with isotropic surface tension numerically
and found that the angle $\phi< 0$  for any positive $\sigma$.

\begin{figure}
\begin{center}
\epsfig{file=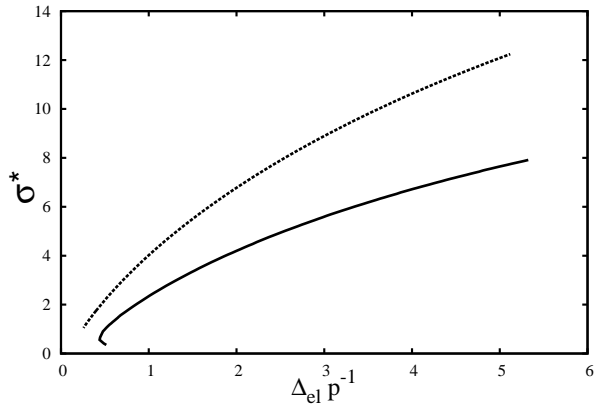, angle=-90, width=8cm}
\caption{Stability parameter $\sigma^{\ast}$ versus 
$\Delta_{el}/p$ for two values of  $\phi$: the dashed line corresponds to $\phi=0$, 
and the solid line corresponds to $\phi=\pi/6$.
   }
\label{alpha}
\end{center}
\end{figure}

Now, we discuss the numerical results obtained by the solution of 
Eq. (\ref{full_nonlin_main})  in the spirit of Ref. \cite{meiron}. 
 In the important regime of  small Peclet numbers, the  eigenvalue 
$\sigma=\sigma^{\ast}(\phi^{\ast},\Delta_{el},p)$ depends only  on the ratio   
$\Delta_{el}/p$ for a fixed angle $\phi$. While the strength of the elastic effects is 
assumed to be small, $\Delta_{el}\ll 1$, the control parameter $\Delta_{el}/p$ can be 
varied in a wide region in the limit of small $p$.  
The eigenvalue $\sigma^{\ast}$ as a function of $\Delta_{el}/p$ for two  values of 
the  angle, $\phi=0$ and $\phi=\pi/6$, is shown in Fig.\ref{alpha}. 
The situation with $\phi\approx 0$ is realized if $\gamma_b\ll\gamma$, 
while $\phi\approx\pi/6$ corresponds to $\gamma_b\approx \gamma$. 
The Poisson ratio was fixed to $\nu=1/3$. 
%%%%%figure%%%%%%%%%%%%%%%%%%%%%%%%%%%%%%

%%%%%%%%%%%%%%%%%%%%%%%%%%%%%%%figure%%%%%%%%%%%%%%%%%%%%%%%%%%%%%%%%%%%%%
The most remarkable feature of these results is  that we do find dendritic 
solutions for the isotropic surface tension in the presence of the elastic effects.
 In this sense, the elastic effects serve as a new selection mechanism. 
 We note that $\sigma^{\ast}$ becomes large for large values of $\Delta_{el}/p$, while in the classical dendritic growth $\sigma^{\ast}$ is always small, being controlled by tiny anisotropy effects. Thus, the growth velocity,
 $\upsilon=2D\sigma^{\ast}p^2/d_0$, can be much larger due to elastic effects, compared to the classical dendritic growth.

In the case $\phi=\pi/6$, the solution exists only beyond some critical value 
of the control parameter, $\Delta_{el}/p$. The lower branch of this solution has a 
negative curvature near the triple junction and it is presumably unstable. 
The general structure of the theory suggests that in the case $\phi=0$, the curve 
should start from the origin ($\sigma=0,\Delta_{el}/p=0$) which is a singular point 
of the problem. However, numerics becomes very difficult in the vicinity of this point. 

We have also performed several runs for the single crystal growth ($\theta=0$ in 
Eq. (\ref{strain1})). As in the classical dendritic growth, we have not found solutions 
with a smooth tip, $\phi=0$. Moreover, for  negative values of $\phi$, 
where solutions exist and  correspond to the growth along a grain boundary in the mother 
$\a$ phase \cite{pilip07}, 
the selected stability parameter $\sigma^{\ast}$ is a decreasing function of 
$\Delta_{el}/p$. This is in strong contrast with the results for the bicrystal growth.
    
We hope that our results will stimulate new experimental and theoretical investigations 
in this interesting field. On the theoretical side, a challenging problem would be the 
investigation of the doublon patterns in the case of bicrystal growth. 
With this scenario,  two crystals of the $\b$ phase would be separated by a thin 
film of the mother $\a$ phase and not by the twin boundary. This pattern is a 
competitive structure at least in the range of the small control parameter 
$\Delta_{el}/p$, where the dendritic pattern with the twin boundary does not exist. 

This work has been supported in part by the Deutsche  Forschungsgemeinschaft under 
Grant SSP 1296 and by the German-Israeli Foundation.   

\end{document}